\documentclass[reprint,journal=jctc]{achemso}

\usepackage{graphicx}
\usepackage{color}
\usepackage{amsmath}
\usepackage[english]{babel}
\usepackage[nolist]{acronym}
\usepackage{physics}
\usepackage[sort&compress]{natbib}

\newcommand{\D}{\mathrm{d}}
\newcommand{\I}{\mathrm{i}}
\newcommand{\E}{\mathrm{e}}
\newcommand{\FeCO}{[Fe(CO)$_5$]$^{0}$}
\newcommand{\FeHO}{[Fe(H$_2$O)$_6$]$^{2+}$}

\newcommand{\TiO}{[TiO$_6$]$^{8-}$}
\newcommand{\Hmol}{H$_2$}
\newcommand{\hcci}{HCCI} 
\newcommand{\benz}{C$_6$H$_6$} 
\newcommand{\RD}{\texttt{RhoDyn}}
\newcommand{\molcas}{\texttt{OpenMOLCAS}}

\title{RhoDyn: a \acs{TD-RASCI} framework to study ultrafast electron dynamics in molecules} 

\author{Vladislav Kochetov}
\affiliation{Institut f\"{u}r Physik, Universit\"{a}t Rostock, A.-Einstein-Strasse 23-24, 18059 Rostock, Germany}

\author{Sergey I. Bokarev}
\email{sergey.bokarev@uni-rostock.de}
\alsoaffiliation{Institut f\"{u}r Physik, Universit\"{a}t Rostock, A.-Einstein-Strasse 23-24, 18059 Rostock, Germany}

\SectionNumbersOn

\begin{document}


  \begin{abstract}
	This article presents the program module \RD\ as part of the \molcas\ project intended to study ultrafast electron dynamics within the density-matrix-based time-dependent restricted active space configuration interaction framework (\acs{TD-RASCI}). 
	The formalism allows for the treatment of spin-orbit coupling effects, accounts for nuclear vibrations in the form of a vibrational heat-bath, and naturally incorporates (auto)ionization effects.
	Apart from describing the theory behind and the program workflow, the paper also contains examples of its application to the simulations of the linear L$_{2,3}$ absorption spectra of titanium complex, high harmonic generation in the hydrogen molecule, ultrafast charge migration in benzene and iodoacetylene, and spin-flip dynamics in the core-excited states of iron complexes.

  \end{abstract}

  \maketitle

\begin{acronym}
	\acro{ADC}{Algebraic Diagramatic Construction}
	\acro{CAS}{Complete Active Space}
	\acro{CASPT2}{Complete Active Space Second Order Perturbation Theory}
	\acro{CASSCF}{Complete Active Space Self-Consistent Field}
	\acro{CI}{Configuration Interaction}
	\acro{CM}{Charge Migration}
	\acro{CSF}{Configuration State Function}
	\acro{DO}{Dyson orbital}
	\acro{DKH}{Douglas-Kroll-Hess}
	\acro{DMRG}{Density Matrix Renormalization Group}
	\acro{GAS}{Generalized Active Space}
	\acro{HHG}{High Harmonics Generation}
	\acro{MCTDH}{Multi-Configurational Time-Dependent Hartree}
	\acro{MO}{Molecular Orbital}
	\acro{NL}{Non-Linear}
	\acro{PES}{Photoelectron spectrum}
	\acro{RAS}{Restricted Active Space}
	\acro{RASPT2}{Restricted Active Space Second Order Perturbation Theory}
	\acro{RASSCF}{Restricted Active Space Self--Consistent Field}
	\acro{RASSI}{Restricted Active Space State Interaction}
	\acro{SCF}{Self--Consistent Field}
	\acro{SF}{Spin-Free}
	\acro{SO}{Spin-Orbit}
	\acro{SOC}{Spin-Orbit Coupling}
	\acro{TD-CI}{Time-Dependent Configuration Interaction}
	\acro{TD-GASCI}{Time-Dependent Generalized Active Space Configuration Interaction}
	\acro{TD-MCSCF}{Time-Dependent Multi-Configurational Self-Consistent Field}
	\acro{TD-RASCI}[$\rho$-TD-RASCI]{Density-matrix Based Time-Dependent Restricted Active Space Configuration Interaction}
	\acro{TM}{Transition Metal}
	\acro{TDSE}{Time--Dependent Schr\"{o}dinger Equation}
	\acro{XAS}{X-ray Absorption Spectrum}
	\acro{XES}{X-ray Emission Spectrum}
	\acro{XFEL}{X-ray Free Electron Laser}
\end{acronym}


\section{Introduction}\label{sec:introduction}
  The last decade heralds the gradual change of the ultrafast paradigm in physics and chemistry from the femtosecond to subfemtosecond and even a few tens of attoseconds domain. 
  The fascinating growth in the number of studies of the ultrafast phenomena owes to establishing new sources such as \acp{XFEL}~\cite{McNeil_NP_2010, Grguras_NP_2012, Maroju_N_2020, Serkez_JO_2018} and \ac{HHG} setups~\cite{Hentschel_N_2001, Gaumnitz_OE_2017} which give access to dynamics at electronic timescales.~\cite{Schultz2014, Young_JPB_2018, Baykusheva2020, Saalfrank_AiQC_2020}
  State-of-the-art \acp{XFEL} allows studying processes with extremely intense ultra-short pulses enabling studies of multiple ionization and radiation damage.~\cite{Doumy_PRL_2011, Rudek_NP_2012}.
  \ac{HHG}, in turn, gives unparalleled pulse duration of several tens of attoseconds~\cite{Gaumnitz_OE_2017} and thus enables unprecedented experiments on electronic structure, such as imaging of molecular orbitals,~\cite{Niikura_N_2002, Itatani_N_2004, Villeneuve_S_2017} attosecond interferometry,~\cite{Smirnova_N_2009, Smirnova_PNAS_2009} 
  measuring phases of photoionization amplitudes,~\cite{Richter_PCCP_2019, You_PRX_2020} and others.
  Moreover, one of the most intriguing phenomena in ultrafast physics and chemistry, \ac{CM},~\cite{Kuleff_JPB_2014, Nisoli_CR_2017} has already got impressive experimental evidence.~\cite{Calegari_S_2014, Kraus_S_2015, Worner_SD_2017}

  From the viewpoint of theory, quantum chemistry packages provide an accurate static electronic structure. 
  However, to meet the challenges of time and improve the interpretation of the experimental data, one needs to predict the sub-femtosecond electron dynamics in molecules and extended systems.
  That is why an extension of quantum chemistry to the time domain is warranted.
  
  There are plenty of methods dealing with ultrafast phenomena occurring on the attosecond to few-femtosecond timescale.~\cite{Alvermann_NJP_2012, Moskalenko_PR_2017, Goings_WIRCMS_2018, Li_CR_2020}
  For instance, versatile methods are formulated in the framework of \ac{ADC},~\cite{Schirmer_PRA_1983, Kuleff_JCP_2005} the coupled-cluster family of approaches,~\cite{Hoodbhoy_PRC_1978, Kvaal_JCP_2012, Nascimento_JPCL_2017} and time-dependent density functional theory.~\cite{Lopata_JCTC_2012}
  The \ac{MCTDH}(F) method~\cite{Meyer_CPL_1990, Beck_PR_2000, Nest_JCP_2013, Despre_PRL_2018, Lode_RMP_2020} should be named among the multi-configurational wave-function techniques allowing to study electron dynamics in real space and time.
  In energy representation (state basis), a traditional approach to electron dynamics is \ac{TD-CI}~\cite{Krause_JCP_2005,Klamroth_PRB_2003, Greenman_PRA_2010,Wang_PRL_2017} or, more generally, the \ac{TD-MCSCF}~\cite{Sato_PRA_2013,Miyagi_PRA_2013} approach. 
  With increasing molecular size or considering deeper-lying core orbitals, it is crucial to decrease computational cost, where the concepts of \ac{RAS}~\cite{Miyagi_PRA_2013} and \ac{GAS}~\cite{Bauch_PRA_2014} help reduce the number of electronic configurations.
  Another recently implemented method allowing for larger active spaces is time-dependent \acl{DMRG} (TD-DMRG).~\cite{Haegeman_PRB_2016,Baiardi_JCTC_2019,Frahm_JCTC_2019,Baiardi_JCTC_2021}
  
  
  The methods mentioned above are based on the \ac{CI} wave function decomposition and the subsequent propagation of expansion coefficients.
  However, it is often necessary to describe open quantum systems in a more general way. 
  For instance, the density matrix formulation offers some advantages. 
  It allows for an implicit inclusion of environmental effects such as dephasing and energy relaxation and natural incorporation of (auto)ionization~\cite{Tremblay_JCP_2008, Tremblay_JCP_2011} phenomena.
  Moreover, non-linear spectroscopies~\cite{Gelin_JCP_2003, Mukamel_ARPC_2013,  Ando_JACS_2014, Zhang_TCC_2015} are usually formulated in terms of perturbation expansion of the density matrix.
  Thus, a wider range of phenomena can be investigated within the density-matrix-based approach.

  In this communication, we present an implementation of the \ac{TD-RASCI} method into the \molcas\ program package~\cite{FernandezGalvan_JCTC_2019} within a program module called \RD.
  The primary goal of this module is to provide access to the various aspects of the ultrafast electron dynamics in molecules, including the influence of \ac{SOC} effects, population and phase relaxation due to the environment, and (auto)ionization. 
  The manuscript is organized as follows:
  We start from the theory underlying the method in Section~\ref{sec:methodology}.
  Section~\ref{sec:workflow} describes the program's functionality.
  Further, in Section~\ref{sec:examples}, we present some exemplary applications to the \ac{HHG} spectrum of the hydrogen molecule, charge-migration dynamics in benzene and iodoacetylene, and spin-dynamics in the core-excited states of transition metal complexes.
  Section \ref{sec:conclusion} summarizes the peculiarities of the \RD\ module and provides an outlook of future developments.

\section{Methodology}\label{sec:methodology}

  \begin{figure*}
	\includegraphics[width=0.85\textwidth]{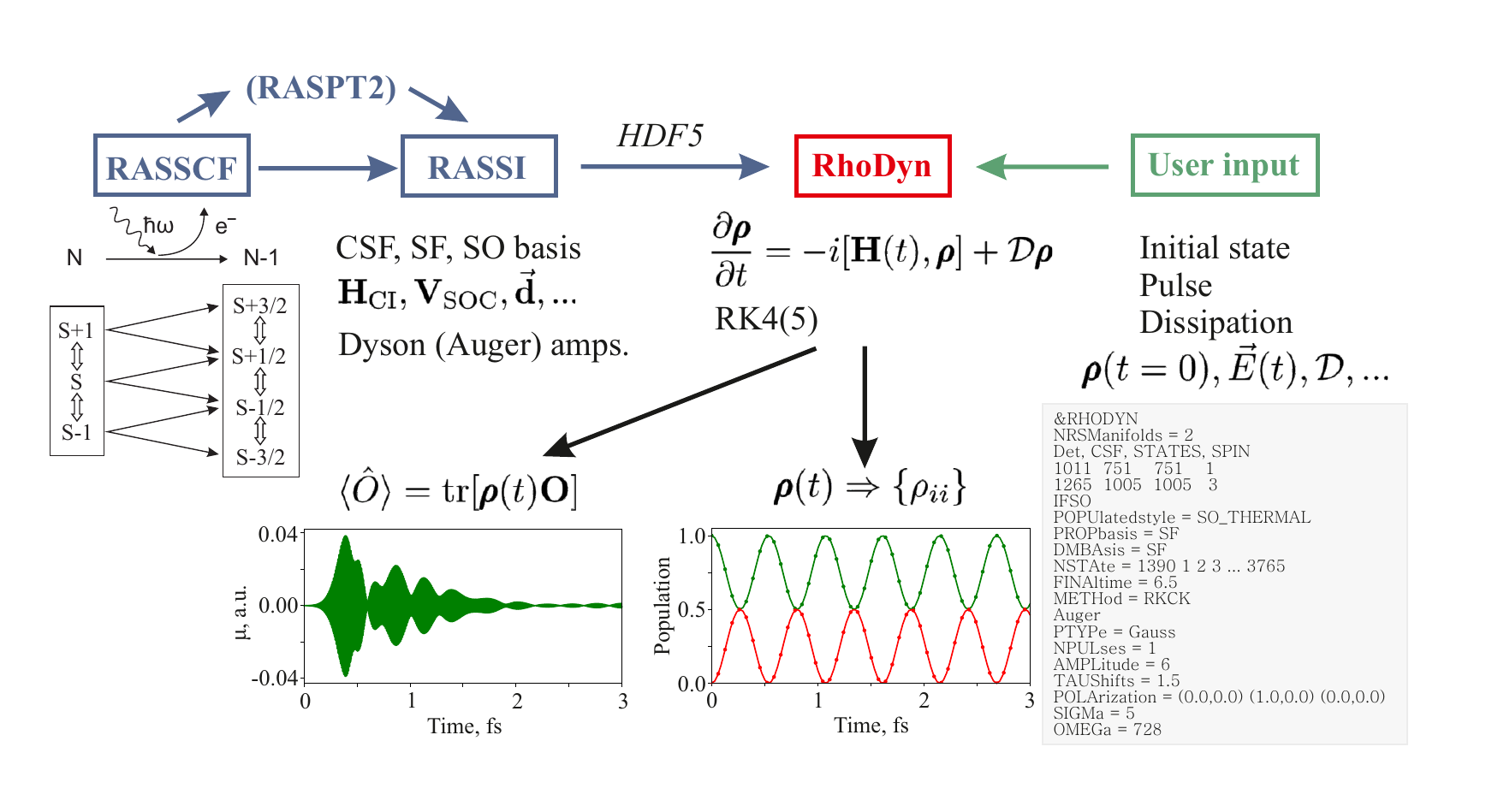}
	\caption{\label{fig:scheme} Scheme of the workflow and dependencies of the \RD\ module on other parts of the \texttt{OpenMolcas} package.}
	\end{figure*}

  \ac{TD-RASCI} method~\cite{Wang_MP_2017} is implemented as a core feature of the \RD\ program module.
  It is intended to study purely electronic dynamics when nuclear motion does not play an important role, completely altering the dynamics.
  Such an approach seems to be especially useful to study core-state dynamics since electron dynamics are to a large extent isolated from nuclear effects owing to the characteristic timescale of core electron's motion and the ultrashort lifetime of the core hole not exceeding few fs. 
  To still be able to take the influence of the energy and phase relaxation due to vibronic interactions into account, the \RD\ module allows employing the electronic system--vibrational bath partitioning; for details, see Ref.~\citenum{Kochetov_JCP_2020}.
  In such an approach, the dynamics of an open system are described via its reduced density operator $\hat \rho$ following the Liouville-von Neumann equation 
  \begin{equation}\label{eq:liouville}
  	\frac{\partial}{\partial{t}}\hat{\rho}=-i[\hat{H},\hat{\rho}]-\mathcal{D}\hat{\rho} \, ,
  \end{equation}
  with a dissipation superoperator $\mathcal{D}$.
  Note that here and below atomic units are used unless stated otherwise.
  The \RD\ module is inherently interfaced with other core programs of the \molcas\ package, as shown schematically in Fig.~\ref{fig:scheme}.
  The matrix (tensor) forms of the $\hat \rho$, $\hat H$, and $\mathcal{D}$ operators are written in energy representation using the eigenstates of some zero-order Hamiltonian $\hat H_0$.
  The program allows for a flexible choice of the basis.
  On a basic level, the basis of \acp{CSF} is utilized.
  In this latter basis, the Hamiltonian matrix takes the form
  \begin{equation} \label{eq:hamiltonian}
  	\mathbf{H}(t)= \mathbf{H}_{\rm CI}+\mathbf{V}_{\rm SOC}+\mathbf{U}_{\rm ext}(t) \, .
  \end{equation}
  Here, $\mathbf{H}_{\rm CI}$, $\mathbf{V}_{\rm SOC}$, and $\mathbf{U}_{\rm ext}(t)$ are the time-independent \ac{CI} Hamiltonian responsible for electron correlation effects, \ac{SO} interaction, and time-dependent external potential, e.g., due to interaction with the light field, respectively.
  The study of electron-correlation-driven dynamics can be conveniently studied in this basis.
  However, for more complicated processes, the investigator might prefer to take the eigenfunctions of the $\hat H_0=\hat H_{\rm CI}$ or $\hat H_0=\hat H_{\rm CI}+\hat V_{\rm SOC}$ operators as a basis.
  We will call them \ac{SF} and \ac{SO} states in what follows.
  In its simplest form, the light-matter interaction term is represented as semi-classical dipole coupling $\mathbf{U}_{\rm ext}(t) = -\vec{\mathbf{d}} \cdot \vec{E}(t)$, where $\vec{\mathbf{d}}$ is a transition dipole tensor written in one of the bases mentioned above, and $\vec{E}(t)$ is an external electric field, see Sec.~\ref{sec:workflow}.	

  The quantities necessary for the propagation, $\mathbf{H}_{\rm CI}$, $\mathbf{V}_{\rm SOC}$, $\vec{\mathbf{d}}$, and the transformation matrices between \ac{CSF}, \ac{SF}, or \ac{SO} bases, are transferred from the \texttt{RASSCF} and \texttt{RASSI} modules of \molcas, see Fig.~\ref{fig:scheme}.
  In \RD, the user needs to supply the form of the light field $\vec E(t)$ and the dissipation tensor $\mathcal{D}$, which can be obtained numerically as described in detail elsewhere \cite{Kochetov_JCP_2020} or take a simple parametrized phenomenological form.
  	

 Propagation of the density matrix according to Eq.~\eqref{eq:liouville} is performed with the adaptive Runge-Kutta-Cash-Karp method \cite{Cash_ATMS_1990,Press1996} of the 4(5) order of accuracy or with the fourth-order Runge-Kutta with a fixed timestep. 
 In many cases, this method suffices as it approximates the full exponential propagator sufficiently accurately to produce the same results.~\cite{Schriber_JCP_2019,Frahm_JCTC_2019}
  	
  The main output of the \RD\ consists of the time-dependent reduced density matrix $\mathbf{\rho}(t)$, which can be printed out with any convenient timestep.
  Its diagonal provides occupation numbers of the basis states.
  More importantly, this matrix can be used to compute the expectation value of any operator $\hat O$ whose matrix is written in the same basis:	
  %

 \begin{equation}\label{eq:observable}
  	\langle \hat{O} \rangle = \mathrm{tr}[\pmb{\rho}(t) \mathbf{O}]\, .
 \end{equation}
In this respect, the most prominent example is the dipole moment $\langle\hat\mu\rangle(t)$, as it provides access to linear and non-linear spectra of the system, see, e.g., Sections~\ref{subsec:ti_xas} and~\ref{subsec:hhg}.


\section{Computational workflow}\label{sec:workflow}
  
  The \RD\ module vastly relies on the infrastructure of the \molcas\ package.
  The workflow of a dynamical calculation, including the example of input, is illustrated in Fig.~\ref{fig:scheme}.
  First, one needs to compute all the wave functions with \ac{CASSCF} or \ac{RASSCF} methods for all state-manifolds which are relevant for the dynamics with the \texttt{RASSCF} module; these can be states of different multiplicity coupled via \ac{SOC} or states with a different number of electrons if photoionization, autoionization, or electron attachment are considered.
  One might wish to employ a \ac{CASPT2} or \ac{RASPT2} energy correction to include dynamic correlation.
  Finally, the \texttt{RASSI} module implementing the \ac{RASSI} method~\cite{Malmqvist_CPL_2002} provides $\mathbf{H}_{\rm CI}$ and $\mathbf{V}_{\rm SOC}$, transition dipole matrix $\vec{\mathbf{d}}$, entering the Hamiltonian Eq.~\eqref{eq:hamiltonian}, in any convenient basis of \acp{CSF}, \ac{SF}, or \ac{SO} states and the transformation matrices between the bases.
  The communication of the data from the respective modules to \RD\ is done via the \texttt{HDF5} interface.~\cite{hdfgroup}
  

  	The user needs to supply the \RD\ with the initial density matrix, which can be, for instance, represented as the thermal ensemble in equilibrium 
  	\begin{equation}
  	\rho_{ij}(0)=\delta_{ij}\exp(-E_i/(kT))\, .
  	\end{equation}  
  	However, it can be generally constructed from any state-vector $\ket{\Psi}$ in the respective basis $\{\ket{\Phi_i}\}$ as $\pmb{\rho}(0)=\sum_{ij} \braket{\Phi_i}{\Psi}\braket{\Psi}{\Phi_j}$ and be read in a matrix form from a separate file.
    
    In the absence of static contribution, the electric field can be derived from the vector potential as $\vec{E}(t)=-\partial \vec{A}(t)/\partial t$; thus, both the oscillatory function and the pulse envelope need to be differentiated.
    It gives rise to two terms, e.g., for a Gaussian-shaped light pulse with vector potential $\vec{A}(t)={A}{\Omega}^{-1}\vec{e}\exp{-(t-t_0)^2/(2\sigma)^2}\cos{(\Omega t + \varphi_0)}$ {(note that normalization factor of the envelope is included in the amplitude $A$)}, the electric field reads
    \begin{align}\label{eq:gaussian}
    \vec{E}(t)&=A\vec{e}\exp{-(t-t_0)^2/(2\sigma^2)}\sin{(\Omega t + \varphi_0)} \nonumber\\
    &+\frac{A\vec{e}(t-t_0)}{\Omega\sigma^2}\exp{-(t-t_0)^2/(2\sigma)^2}\cos{(\Omega t + \varphi_0)}\, .
    \end{align}
    Here, $A$, $\vec e$, $t_0$, and $\Omega$ are the amplitude, polarization, time-center of the envelope, and carrier frequency.
    The second correction term ensures that the integral of the electric field over the entire pulse vanishes $\int_{-\infty}^{\infty} \vec{E}(t) dt = 0$.~\cite{Paramonov_JPCA_2012}
    However, it should be notable only for small $\Omega$, e.g., valence excitations, when the carrier-envelope phase $\varphi_0$ matters.
    For core excitations and pulse durations of more than 200\,as, this term is of minor importance.
  	
  	In \RD, the user can choose between different options for pulse forms, such as Gaussians and more localized $\sin^n(\pi(t-t_0)/(2\sigma))$ or $\cos^n(\pi(t-t_0)/(2\sigma))\quad (n=2,...)$.
  	There is also a possibility to select a linearly chirped pulse \cite{Schultz2014} with {$\Omega(t)=\Omega_0+a(t-t_0)$.}
  	Apart from a single pulse, one can choose a sequence with individual polarization, intensity, duration, time shift, and carrier frequency.
  	Thus, it enables calculations of the non-linear spectra. 
  	Currently, in \RD, there are no tools to perform orientational averaging, but this can be done at the postprocessing stage.

  The Redfield tensor $\mathcal{D}$ in Eq.~\eqref{eq:liouville} accounts for the coupling to the vibrational bath.~\cite{May2011, Blum2012}
  The decay rates must be calculated separately; see, e.g., Ref.~\citenum{Kochetov_JCP_2020}, and \RD\ reads them in a matrix form.
  The user can also complement the diagonal of the Hamiltonian by imaginary numbers to implicitly account for some other decay channels.
  In this case, the propagation is non--norm-conserving and $\tr \pmb{\rho} < 1$.
  
%


\section{Exemplary applications}\label{sec:examples}

  Below we present three different applications of the implemented methodology: 
  i) using the calculated time-dependent dipole moment to obtain the linear \ac{XAS} of \TiO\ cluster and the high harmonic generation spectrum of the H$_2$ molecule triggered by a strong-field IR laser pulse, 
  ii) the charge migration in the benzene and iodoacetylene molecules caused by sudden ionization and short UV pulse, and 
  iii) the ultrafast spin-flip dynamics in the \FeHO\ and \FeCO\ complexes in the core-excited states triggered by an ultrashort X-ray pulse.
  Possible applications are, of course, not limited to these types of processes and may include studies of the multiple ionization and non-linear spectroscopies as will be detailed elsewhere.

  \subsection{Linear XAS}\label{subsec:ti_xas}

  The time-dependent dipole moment $\vec{\mu}(t)$ can be used for computing multi-time correlation functions and response functions of different orders to simulate and understand non-linear spectra.~\cite{Mukamel1999}
  The simplest example is a linear absorption spectrum which can be obtained as a Fourier-transform of the $\vec \mu (t)$:
  \begin{equation}
  \alpha(\omega) = \frac{1}{2\pi}\Im \left[ \int_0^{t_f} \D t\ \E^{\I \omega t} \vec{\mu}(t) \cdot \vec{e}\ W(t) \right].
  \end{equation}
  Here, the oscillations of the $\vec\mu$ are initiated by the incoming pulse with polarization $\vec e$, $t_f$ is the length of propagation, and $W(t)$ is a window function used to filter out noise. 
  
\begin{figure}
  	\includegraphics[width=0.5\linewidth]{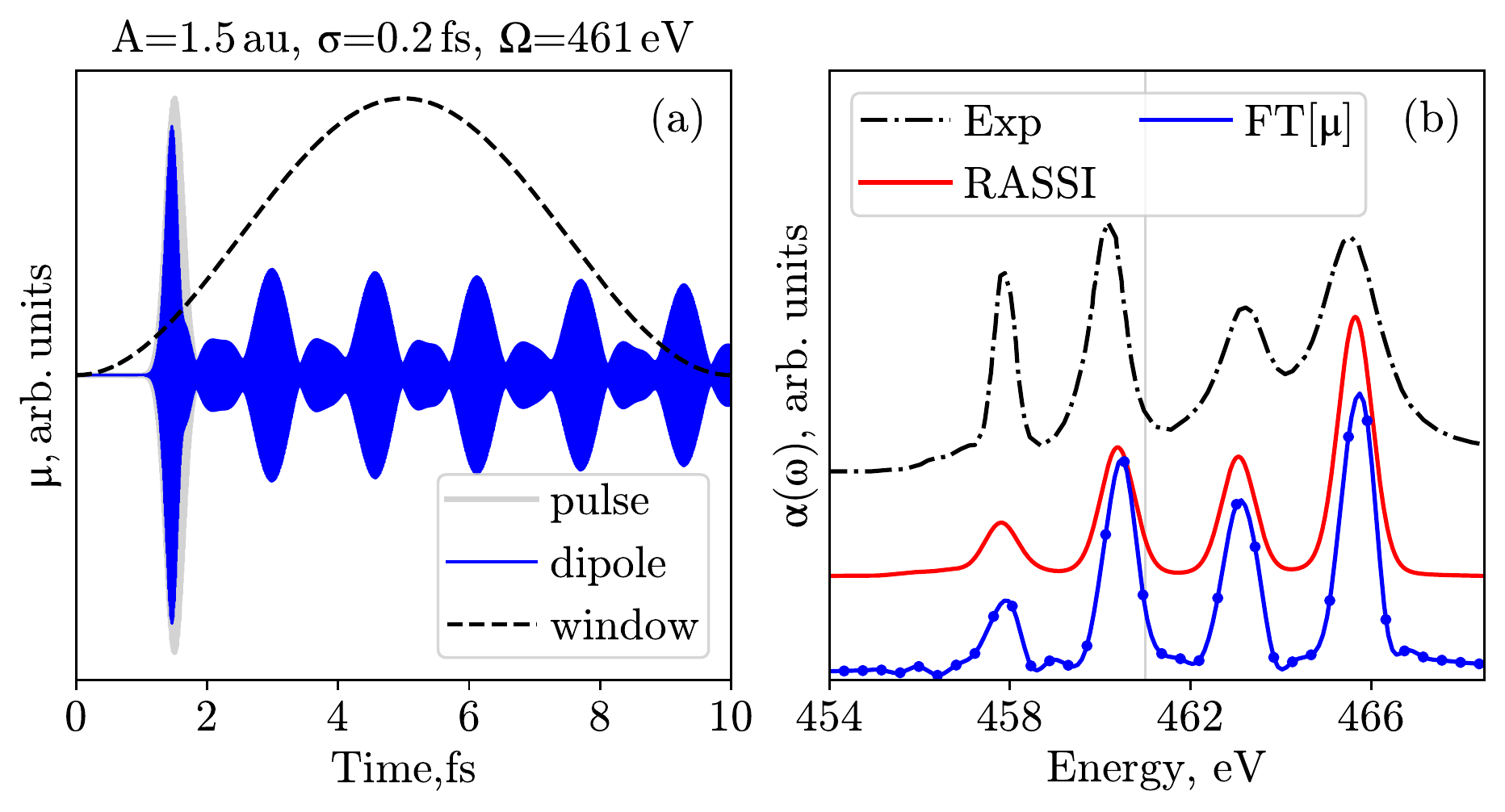}
  	\caption{\label{fig:ti}  
  		a) ultrashort X-ray pulse (gray shaded area) with characteristics given above the panel, time-evolution of $\vec\mu(t)$ (blue), and Hann window function (dashed) used to simulate the \ac{XAS} of the \TiO\ cluster.
  		b) comparison of the Fourier transform of $\vec{\mu}(t)$ (FT[$\mu$], blue) and the static spectrum obtained from \texttt{RASSI} (red) with experiment.~\cite{Woicik_PRB_2007}
  		The thin vertical line marks the carrier frequency of the excitation pulse.}
  \end{figure}

  An example is given in Fig.~\ref{fig:ti}, displaying the L$_{2,3}$-edge \ac{XAS} of the \TiO\ cluster mimicking the bulk TiO$_2$.
  The calculation has been performed on the \ac{RASSCF} level with the ANO-RCC-VTZP basis, including three 2p and five 3d orbitals of titanium atom in the RAS1 and RAS3 spaces allowing for one hole/electron, respectively.
  Thus, it corresponds to the \ac{CI} singles level of theory.
  The spectrum is globally shifted by 8.1\,eV for better comparison with the experiment.
  For further details of calculations, see Ref.~\citenum{Kochetov_JCP_2020}.
  Here, the propagation length of $t_f$=10\,fs has been used. 
  We employ the Hann window {$W(t)=\sin^2(\pi n/N)$, where $n$ is the counting number of $N$ sampling points equal to the number of time points.} 
  The pulse, dipole response, and the window function are shown in panel a) of Fig.~\ref{fig:ti}. 
  Panel b) displays the steady-state spectrum corresponding to time-independent energies and transition dipole moments obtained from the \ac{RASSCF}/\ac{RASSI} calculation and the Fourier-transformed spectrum.
  Although the time-dependent procedure is redundant with the time-independent one, both results agree reasonably, representing an important consistency check.
  %

  \subsection{High-harmonic generation}\label{subsec:hhg}
  \ac{HHG} is a highly non-linear optical effect observed for atomic and molecular gases as well as for solids.~\cite{Ghimire_NP_2011, Ndabashimiye_N_2016, Lakhotia_N_2020}
  As a result of the interaction of a high-intensity light pulse having carrier frequency $\Omega$ with the target system,
  the emission of higher harmonics with frequencies $N\Omega$ occurs; $N$ is odd for the bright harmonics for the isotropic systems. 
  Such a high-energy spectrum is due to the field-driven recombination of the accelerated electron with the ionized target.
  
  \ac{HHG} spectra can be calculated in the length form as a Fourier-transform of the dipole moment's response to the incoming radiation with polarization $\vec{e}$~\cite{Saalfrank_AiQC_2020}
  \begin{equation}\label{eq:hhg_main}
  I(\omega) = A \omega^4\left| \int_{0}^{t_f} \D t \E ^{- \I \omega t} \vec{\mu}(t) \cdot \vec{e}\ W(t) \right| ^2 \,;
  \end{equation} 

  \begin{figure}[tb]
  	\includegraphics[width=0.5\linewidth]{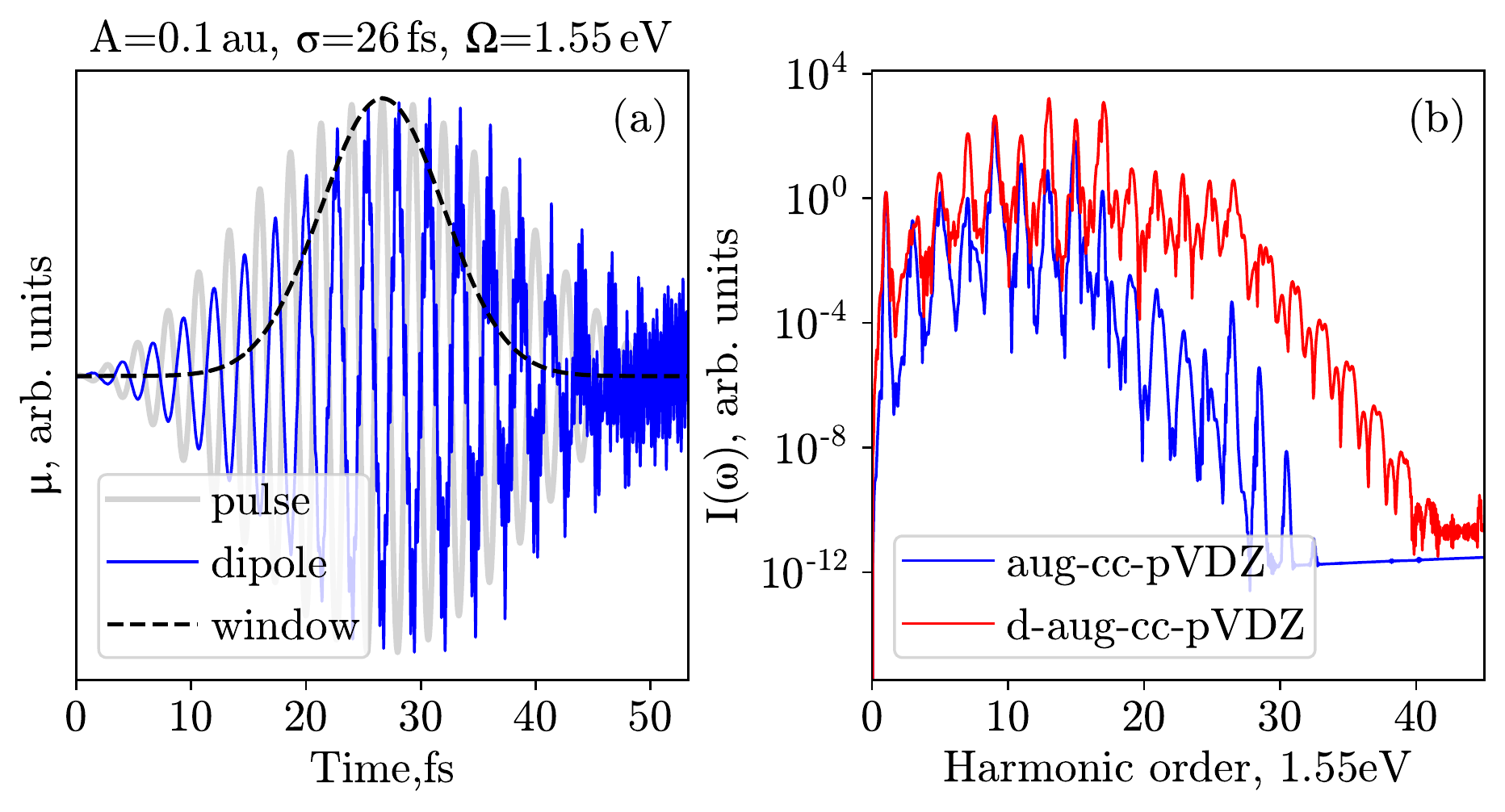}
  	\caption{\label{fig:hhg_h2} (a) Response of the \Hmol\ dipole moment (blue line) to the IR short and intense laser pulse (gray line) for the aug-cc-pVDZ basis. 
  	The filtering window function in Eq.~\eqref{eq:hhg_main} is also depicted with a dashed line. 
  	(b) The resulting \ac{HHG} spectrum of \Hmol\ molecule for the two basis sets.
  	}
  \end{figure}
  
  The \ac{HHG} spectrum has been calculated for a prototypical example of the \Hmol\ molecule.
  Two different basis sets, aug-cc-pVDZ \cite{Kendall_JCP_1992} and d-aug-cc-pVDZ {\cite{Pritchard_JCIM_2019}}, supporting a set of diffuse functions, have been used.
  {The active space comprised all unoccupied orbitals (17 and 25 for both bases, respectively) in the RAS3 space with the only occupied orbital placed in the RAS2 space; further, only single excitations have been allowed to RAS3.
  This setup gave a total of 18 and 26 singlet states for both bases, respectively.}
  Pulse characteristics have been chosen to represent the typical experimental pulse as an output of a Ti-Sapphire laser with $A=3.5 \cdot 10^{15}\,\text{W/cm}^2$ (0.1\,a.u.), $\hbar\Omega$ = 1.55 eV (800 nm), $2\sigma$ = 53.3 fs (including 20 optical cycles), $t_0=0$\,fs, $t_{\rm final} = 2\sigma$, see Fig.~\ref{fig:hhg_h2}a), {and Keldysh parameter $\gamma={\omega\sqrt{2I_p}}/{E_{\rm max}}=1.13$}.
  Pulse envelope corresponds to the $\cos^2$ function.
  {Gaussian window function with the dispersion of 10 fs}, see Fig.~\ref{fig:hhg_h2}a), was applied to time-dependent dipole moment before Fourier transformation.
 
  The two resulting \ac{HHG} for two bases can be seen in Fig.~\ref{fig:hhg_h2}b).
  For the more compact aug-cc-pVDZ basis, the cutoff frequency is observed around the 17th harmonic, whereas for the more diffuse d-aug-cc-pVDZ basis, it shifts to about the 25th harmonic, which demonstrates the importance of taking enough localized Gaussian functions to discretize the continuum relevant for the \ac{HHG} process.
  In comparison with previous studies~\cite{White_MP_2016} and available experimental data~\cite{Mizutani_JPBAMOP_2011}, the results are in reasonable agreement, but of course, the adequately chosen basis set, including diffuse functions and Rydberg states, is of great importance here.~\cite{Luppi_JCP_2013, Ulusoy_JCP_2018}
  One should note that ionization losses should be accounted for by absorbing boundaries, for example, by complex absorbing potential or heuristic ionization model in the spirit of Refs.~\citenum{Klinkusch_JCP_2009, Tremblay_JCP_2011}.
  This procedure delivers a smoother spectrum since the artificial scattering at the boundaries of the localized basis set is decreases.

  \subsection{Charge migration}\label{subsec:charge_migration}
  
  \begin{table*}
  	\caption{\label{tab:charge_migration} {Summary of the charge migration simulations in benzene and iodoacetylene molecules.}}
  	\centering
  	\begin{tabular}{l  l  l}
  		& \benz       & \hcci \\
  		\hline
  		Type of charge migration         & 1h/2h1p     & 1h             \\
  		Character of hole density migration   & Breathing   & From I to C$\equiv$C \\
  		Experimental period, fs   & --          & 1.85 fs \cite{Kraus_S_2015}  \\
  		Theoretical period, this work, fs     & 0.98        & 1.95           \\
  		Theoretical period, other works, fs   & 0.75,~\cite{Schriber_JCP_2019} 0.94,~\cite{Despre_JPCL_2015} 0.80~\cite{Baiardi_JCTC_2021}     & {1.83,~\cite{Schriber_JCP_2019} 1.85~\cite{Frahm_JCTC_2019}} \\
  		Number of basis \acp{CSF} (singlets, doublets)& 175, 210    & 2520, {12096\footnote{{This number includes \acp{CSF} with both $\pm1/2$ spin projections.}}}     \\
  		Number of basis \ac{SF}/\ac{SO} states (singlets, doublets)\footnote{The number of basis states which are actually included in the dynamics.} & 175, 210    & 1, 20-800          \\
  		\hline
  	\end{tabular}
  \end{table*}

  Charge migration represents an attosecond to few-femtosecond oscillatory hole dynamics occurring upon ionization of the system when a superposition of several ionic eigenstates is created, e.g., by a broadband laser pulse. 
  This process is often approximated by an instantaneous removal of an electron from a particular \ac{MO}.~\cite{Kuleff_JPB_2014,Breidbach_JCP_2003}
  This effect has been mainly studied theoretically~\cite{Breidbach_JCP_2003, Kuleff_JCP_2005, Mignolet_PRA_2012, Cooper_PRL_2013,  Nisoli_CR_2017, Jia_JPCL_2019}, although recent experimental advancements also address it.~\cite{Calegari_S_2014, Kraus_S_2015, Worner_SD_2017}
  Its main driving force is electron correlation, but non-adiabatic couplings can also drive it.~\cite{Timmers_PRL_2014}
  More specifically, the basic correlation-driven mechanisms can be different.~\cite{Breidbach_JCP_2003,Kuleff_JPB_2014}
  Here we consider the hole mixing for two examples, benzene and iodoacetylene, which exhibit different types of charge migration.
  The former is {a ``satellite'' migration when a superposition of a 1h and the adjacent 2h1p satellite states is created}, and {the latter is a ``pure'' hole mixing when different 1h states are involved}.
  It implies some differences summarized in Table \ref{tab:charge_migration}.
  
  The mechanism of the preparation of the superposition $\Psi_{\rm ion}$ in theoretical simulations can also be different. 
  One can consider: i) the population of a single $N-1$-electron configuration or \ac{CSF}, $\Psi_{\rm ion}=\Phi_f^{N-1}$;
  ii) the direct action of the annihilation operator for a particular orbital on the $N$-electron ground eigenstate of the unionized system $\Psi_{\rm ion}=\mathcal{N} \hat a_p \Psi_g^N = \mathcal N \sum_j C_{gj} \hat a_p \Phi_j^N$, where $\mathcal N$ is a normalization factor, e.g., $\mathcal N=1/\bra{\Psi_g^N}\hat a_p^\dagger \hat a_p^{} \ket{\Psi_g^N}$;~\cite{Schriber_JCP_2019} or
  iii)  the direct action of the light pulse $\vec E(t)$ on the ground state wave function that conditionally can be denoted as $\Psi_{\rm ion}(t)=\exp[\I\hat{\vec {\mu}}\cdot \vec E(t)t] \Psi_g^N$, where $\hat{\vec {\mu}}$ is the dipole operator.
  The first two ways represent a sudden limit and thus are artificial but convenient in theoretical treatment.
  Case iii) is more realistic and can have a direct correspondence to the experiments.
  Finally, the initial density matrix can be generally constructed from the respective ionic \ac{CI} vectors in the basis of \acp{CSF}, obtained according to i) or ii), as $\pmb{\rho}(0)=\vec{C}_{\rm ion}\cdot\vec{C}^\dagger_{\rm ion}$.


  \paragraph{Benzene}
  
  \begin{figure*}
  	\centering
  	\includegraphics[width=0.7\textwidth]{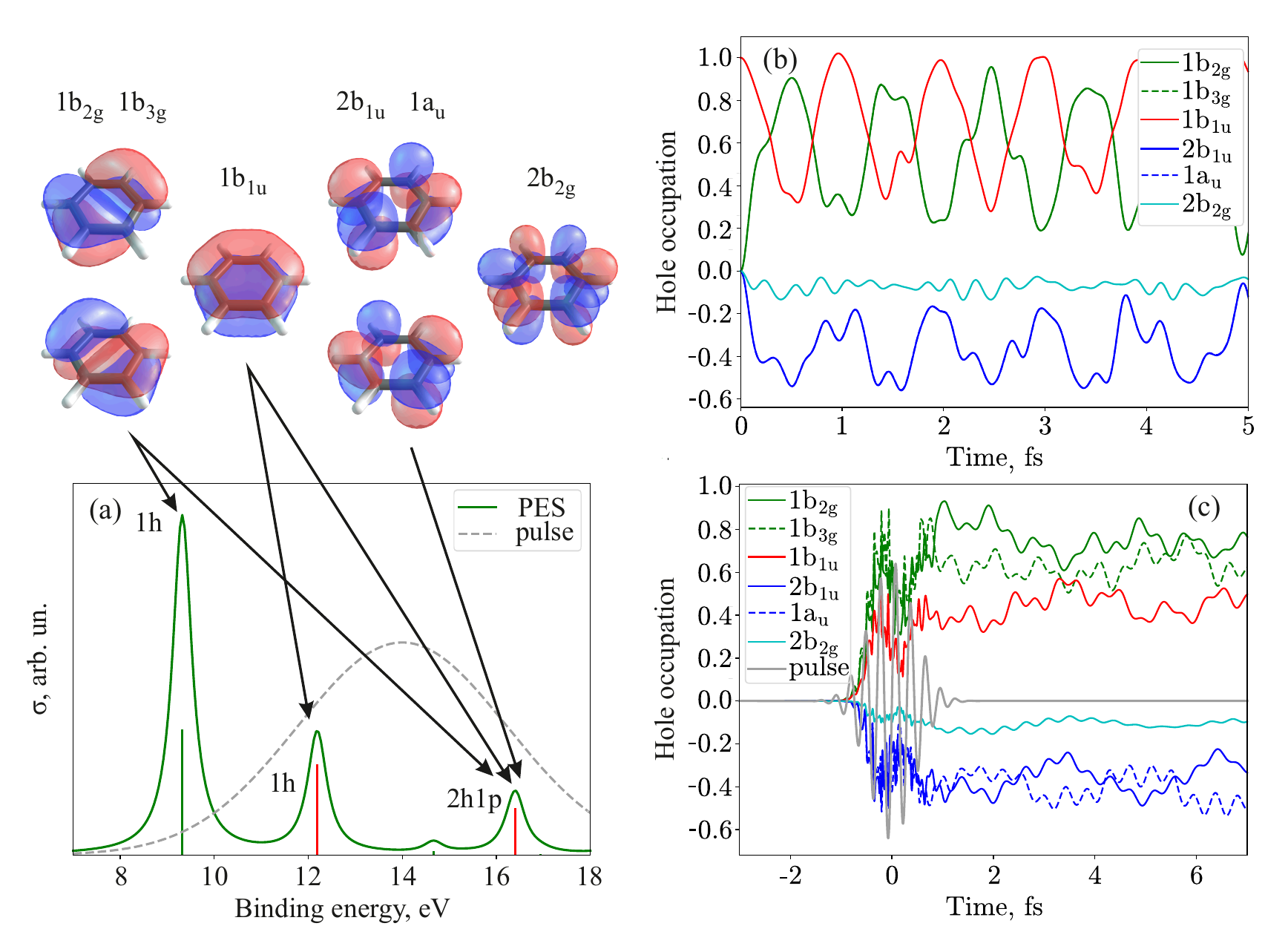}
  	\caption{(a) The \acf{PES} of \benz\ computed at the CASPT2 level using sudden approximation. 
  		The two relevant states/transitions are denoted as red sticks.  
  		The dashed line displays the shape of the pulse used for ionization.
  		The upper part shows the \acp{MO} included in the active space.
  		Arrows show the orbitals which are relevant to the formation of 1h and 2h1p states.
  		(b) Hole occupation dynamics in \benz\ following the population of a single \ac{CSF} with the hole in the $1b_{1u}$ orbital, obtained at the \ac{CASPT2} level of theory. 
  		Negative values correspond to the electron population.
  		(c) Hole occupation dynamics in \benz\ initiated by a pulse with $A=1$\,a.u., $\hbar\Omega=14$\,eV, $\sigma=0.42$\,fs, $\alpha=0.2$, and the pulse polarization parallel to the molecular plane.
  	}
  	\label{fig:benzene}
  \end{figure*}

  We have chosen benzene \benz\ as a convenient example, which is often studied in theoretical works.~\cite{Despre_JPCL_2015, Schriber_JCP_2019, Baiardi_JCTC_2021}
  Charge migration, in this case, consists of hole dynamics following the preparation of the initial state by the sudden removal of an electron from the  $1b_{1u}$ \ac{MO}, Fig.~\ref{fig:benzene}.
  The initial state predominantly represents a superposition of the ``main'' $1b_{1u}$ 1h-state and its 2h1p shake-up satellite due to excitation from the degenerate $1b_{2g}$ and $1b_{3g}$ orbitals to a couple of degenerate unoccupied orbitals $2b_{1u}$ and $1a_{u}$.~\cite{Despre_JPCL_2015}
  Note that orbital notation is given in the largest Abelian subgroup $D_{2h}$ of the full point symmetry group $D_{6h}$ which coincides with the notation of Ref.~\citenum{Schriber_JCP_2019} but differs from that of Ref.~\citenum{Despre_JPCL_2015}.
  It was shown~\cite{Arnold_PRA_2017, Vacher_PRL_2017} that nuclear motion could lead to the loss of coherence at timescales of a few femtoseconds, even for large molecules.
  However, the study of dynamics, including vibrational modes of benzene, resulted in the survival of oscillations~\cite{Despre_JPCL_2015} providing a basis for the clamped nuclei approximation used in this work.
  
  In our calculations active space (6e$^-$,6\acp{MO}) is used containing the complete set of $\pi/\pi^{\ast}$ orbitals.
  The ANO-L-VTZ basis has been employed.
  The dynamics have been computed within the pure $\rho$-TD-CASCI method and also taking the diagonal energy correction due to dynamic electron correlation outside the active space via \ac{CASPT2}.
  {The total number of \acp{CSF} basis functions (equal to the number of accounted \acp{SF}) amounts to 175 singlet and 210 doublet states.}
  Fig.~\ref{fig:benzene} displays the photoelectron spectrum of benzene computed using this setup within the sudden approximation.~\cite{Grell_JCP_2015}
  It has fewer features than the {ADC(3)} spectrum of Despr\'e et al., Ref.~\citenum{Despre_JPCL_2015}, because the ionizations from orbitals outside the active space are not included.
  Nevertheless, it contains all prerequisite states needed to describe the dynamics.
  
  Fig.~\ref{fig:benzene}(b) displays the hole occupation dynamics following the instantaneous occupation of a single \ac{CSF} differing from the main ground state configuration by a hole in the $1b_{1u}$ orbital.
  Such an excitation does not break the symmetry of the molecule, and the occupations of the $1b_{2g}$ and $1b_{3g}$, as well as of $2b_{1u}$ and $1a_u$ (in $D_{2h}$ notation), are the same because these orbitals are degenerate.
  The time evolution of hole occupations 
  was derived from the diagonal elements of the density matrix, $\rho_{ii}{(t)}$, in the \ac{CSF} basis.
  For instance, for an orbital $a$, it reads 
  \begin{equation}
  	n_{{\rm hole},a} (t) = \sum_{i=1}^{N_{\rm CSF}}\rho_{ii}(t)(n_{{\rm GSC},a}-n_{i,a})\, ,
  \end{equation}	
  where $n_{i,a}$ is the occupation number of orbital $a$ for the $i$th \ac{CSF} and $n_{{\rm GSC},a}$ is the respective occupation in the main ground state configuration $(1b_{1u})^2(1b_{2g})^2(1b_{3g})^2(2b_{1u})^0(1a_u)^0(2b_{2g})^0$.
  Hence, the negative hole occupations correspond to the electron occupation.
  
  Panel (b) of Fig.~\ref{fig:benzene} shows the prominent hole dynamics mainly bouncing between $1b_{1u}$ and the $1b_{2g}/1b_{3g}$ pair of orbitals in agreement with previous works.~\cite{Despre_JPCL_2015,Schriber_JCP_2019, Baiardi_JCTC_2021}
  The hole occupation curves demonstrate a bit more wiggles than in the previous works; these features can be assigned to the involvement of the energetically distant ionic states.
  The total of 210 doublet ionic states spans the energy range of {40 eV} in accord with the smallest oscillation period of {around 0.2 fs} seen in the panel (b).
  Calculations at the $\rho$-TD-CASCI level (not shown) are consistent with the adaptive \ac{TD-CI} \cite{Schriber_JCP_2019} and also give the main period of population migration of about 750 as.
  Thus, the additional pair of $\sigma/\sigma^\ast$ included in the active space in Ref.~\citenum{Schriber_JCP_2019} does not play an important role.
  If we apply \ac{CASPT2}, the energy difference between hole-mixed states is calculated more precisely, and then the oscillation period changes to {980\,as} (Fig.\ref{fig:benzene}(b)) and agrees with the results of the ADC(3) method, including a large portion of the dynamic correlation, giving {935}\,as,~\cite{Despre_JPCL_2015} and the TD-DMRG simulations with the (26e$^-$,26\ac{MO}) active space, giving 804\,as.~\cite{Baiardi_JCTC_2021}
  Therefore, the oscillation period is sensitive to the inclusion of the dynamic correlation.
  This conclusion is also supported by a sequence of calculations by Baiardi with increasing active spaces,~\cite{Baiardi_JCTC_2021} leading to the oscillation period's systematic growth.
  
  To address the hole migration dynamics within a more realistic scenario, we employed the \acl{DO} formalism and sudden approximation~\cite{Pickup_CP_1977,Grell_JCP_2015} to populate the ionic state manifold directly from the neutral ground state interacting with the light field.
  Therefore, the transition dipole moment between neutral and ionic states $i$ and $j$ is approximated as $|\vec{{d}_{ij}}|=\alpha||\Phi^{\rm DO}_{ij}||^2$.
  Here, $\Phi^{\rm DO}_{ij}$ is the \ac{DO}, and $\alpha$ is the proportionality factor, which is considered a free parameter governing the degree of the depletion of the ground state.
  Further, we imply that the transition dipole is oriented parallel to the field polarization.
  {For the simulation, we have chosen a Gaussian pulse in the form of the first term in Eq. \eqref{eq:gaussian} linearly polarized parallel to the molecular plane with the following characteristics: $A=1$\,a.u., $\hbar\Omega=14$\,eV, $\sigma=0.42$\,fs;} its form in the frequency domain can be seen in panel (a) of Fig.~\ref{fig:benzene}.
  With this pulse, one predominantly populates the superposition of the target 1h {(12.19\,eV)} and 2h1p {(16.40\,eV)} states, but the other state at {9.32\,eV} binding energy also gets involved.
  
  The results are shown in Fig.~\ref{fig:benzene}(c).
  {The degeneracy of states is lifted by the linearly polarized field leading to uneven occupations of the degenerate orbitals.}
  However, although a prominent hole dynamics is happening, the simulations reveal no characteristic oscillation time in hole occupation due to preparing more complex superposition of the ionic states. 


  \paragraph{Iodoacetylene}
  
  \begin{figure*}
  	\centering
  	\includegraphics[width=0.75\textwidth]{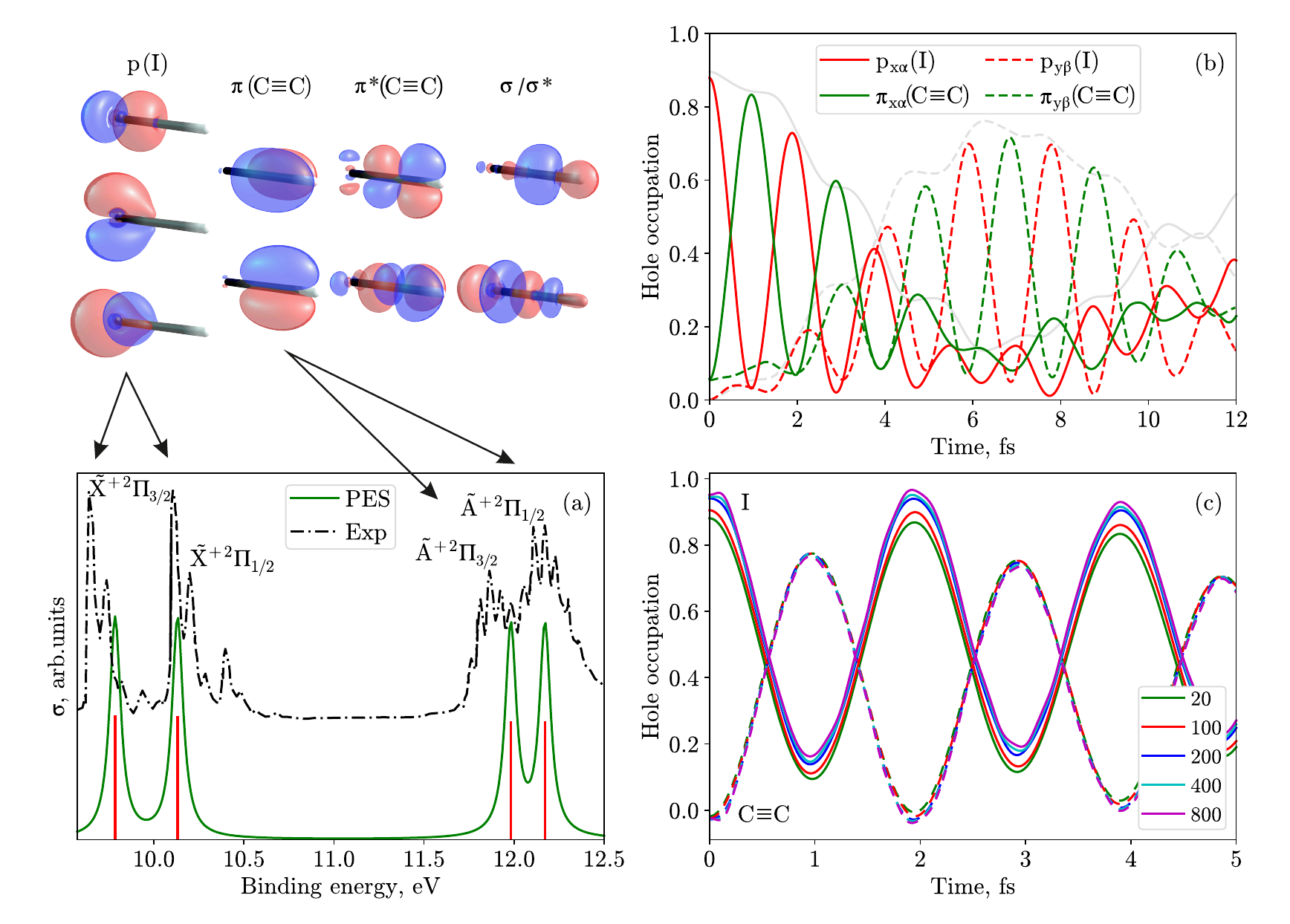}
  	\caption{ (a) \ac{PES} of \hcci\ computed at the \ac{CASPT2} level using sudden approximation.
  		The upper part shows the orbitals in the active space; arrows display the assignment of the respective bands. 
  		The experimental spectrum was digitalized from Ref.~\citenum{Allan_JCS_1976}.
  		(b) Hole occupation dynamics in \hcci\ following the population of a single \ac{CSF} with the hole in the iodine $\mathrm{p_{x\alpha}}$ spin-orbital computed at the \ac{CASPT2} level of theory with 100 states.
  		Occupations of $\alpha$ spin-orbitals are shown in solid lines and those of $\beta$ spin-orbitals in dashed lines.
  		Also shown are the total populations of \acp{CSF} with $S_z= -1/2$ (solid gray line) and +1/2 (dashed gray line).
  		(c) Total hole occupation on the I atom (solid) and C$\equiv$C fragment (dashed) as computed with different numbers of basis \ac{SO} states.}
  	\label{fig:hcci_pes}
  \end{figure*}
   
  In this study, we also focused on the charge migration dynamics in \hcci\ after the instantaneous creation of the hole in the $5p_x(I)$ orbital, which is perpendicular to the molecular axis.
  We also assume that we are able to selectively remove a spin-up electron from this orbital.
  Experimental preparation of such an initial state would require an ensemble of aligned molecules as in Ref.~\citenum{Kraus_S_2015} but in the presence of the magnetic field directed along the axis of the molecules to create a specific superposition of the components of the total angular momentum eigenstates.
  However, here we select this initial state to dissect the effect of the electron correlation responsible for the charge migration and the \ac{SOC} induced dynamics due to the large \ac{SO} constant of iodine.
  
  For \hcci, the active space consists of nine molecular orbitals representing linear combinations of six $2p$-orbitals of carbon atoms and three $5p$-orbitals of iodine. 
  The number of active electrons equals 11 for doublet ionized states and 12 for the initial singlet state; the number of \acp{CSF} is given in Table \ref{tab:charge_migration}.
  ANO-RCC-VTZP basis set with \ac{DKH} Hamiltonian correction~\cite{Douglas_AP_1974} was used for electronic structure calculation to take into account scalar relativistic effects.
  The \ac{CASPT2} energy correction was computed with the imaginary shift of 0.1\,Hartree.
  According to the \ac{PES} presented in Fig.~\ref{fig:hcci_pes}(a), only the states with ionization energies up to 12.5\,eV should be relevant for dynamics amounting to 8 \ac{SO} basis states. 
  Here, we also study the influence of the number of \ac{SO} states, including 20, 100, 200, 400, and 800 states. 
  These choices span the ranges of ionization energies of 16, 21, 24, 28, and 32\,eV, respectively.
  The initial state was prepared by populating the dominant ground-state \ac{CSF} with a removed electron from the $\mathrm{p_{x\alpha}}$ orbital; the dynamics are performed in the basis of \ac{SO} states.

  Photoelectron spectrum, Fig.~\ref{fig:hcci_pes}(a), is obtained in the same way as for benzene molecule.
  Four red sticks are of 1h type and correspond to transitions from the ground state $\tilde{X}\,^{1}\Sigma^{+}$ to $\tilde{X}^{+}\,^{2}\Pi_{3/2}$,
  $\tilde{X}^{+}\,^{2}\Pi_{1/2}$, $\tilde{A}^{+}\,^{2}\Pi_{3/2}$, and $\tilde{A}^{+}\, ^{2}\Pi_{1/2}$ states of the ion.
  The $\tilde{X}^{+}$ pair of bands are primarily associated with the hole in 5p(I) orbitals and $\tilde{A}^{+}$ in  $\mathrm{\pi(C\equiv C)}$ orbitals.
  The experimental spectrum exhibits rich rovibronic structure superimposing on the pure photoionization transitions at 9.71\,eV, 10.11\,eV, 11.87\,eV, 12.12\,eV;~\cite{Allan_JCS_1976} vibrational effects have been not considered in this study. 
  Experimental \ac{SOC} splittings are found to be 0.4\,eV and 0.25\,eV.
  The computed spectrum displays bands with transition energies of 9.78\,eV, 10.13\,eV, 11.98\,eV, 12.17\,eV, respectively, and is in good agreement with the experiment within the accuracy of 0.1\,eV.
  However, the \ac{SO} splittings are predicted slightly lower than in the experiment: 0.35 and 0.19\,eV for $\tilde{X}^{+}$ and $\tilde{A}^{+}$ states.

  The dynamics simulation results are given in Fig.~\ref{fig:hcci_pes}(b), displaying the population of the four mainly involved orbitals.   
  Since the initial density matrix is in the \ac{CSF} basis, we performed a non-orthogonal transformation to the truncated \ac{SO} basis where the number of states is less than the maximum number of \ac{SO} states.
  This transformation leads to a slight loss of the total norm ($\tr[\rho]<1$), as seen in panel (c).
  With the increasing number of \ac{SO} states, the norm is recovered.
  Interestingly, the dynamics character is changing neither qualitatively nor quantitatively since the period of charge oscillations stays the same.
  Only the total hole population on the iodine atom slightly increases.
  The larger number of \ac{SO} basis states, i.e., 800, also introduces slight high-frequency oscillations due to the minute involvement of the energetically-distant eigenstates.
  One can conclude that including more eigenstates than indicated by the photoionization spectrum leads only to a minor improvement in accuracy. 
  
  As seen from panels (b) and (c) of Fig.~\ref{fig:hcci_pes}, the hole migrates from the iodine atom to the C$\equiv$CH fragment with a period of 1.95\,fs. 
  This result is in good agreement with the experimentally found period of 1.85\,fs.~\cite{Kraus_S_2015}
  The initial dynamics involve the \acp{CSF} with $S_z=-1/2$, and thus the hole occupies $\alpha$ spin-orbitals.
  However, due to the notable \ac{SOC} of iodine, the hole also populates $\mathrm{p_{y\beta}(I)}$ and $\mathrm{\pi_{y\beta}(C\equiv C)}$ orbitals, shown with dashed lines.
  The oscillations in the y-oriented $\beta$ manifold are slightly {retarded} compared to the x-oriented $\alpha$ orbitals.  
  Thus, the dynamics correspond to pumping the population from the \ac{CSF}-manifold with $S_z=-1/2$ to $S_z=+1/2$ and back with a period of about 12\,fs as shown with gray lines in panel (b).
  Its timescale agrees with an average \ac{SO} splitting of the $\tilde{X}^{+}$ and $\tilde{A}^{+}$ states of 0.3\,fs.
  All other orbitals stay insignificantly occupied with the summed population of less than 0.1.
 
  The computed period of 1.95\,fs is only slightly larger than in other theoretical works~\cite{Frahm_JCTC_2019,Schriber_JCP_2019} with the notably larger active space, including 36 and 22 active orbitals with 16 active electrons, respectively.
  Again, this fact evidences that some portion of electron correlation, which is essential for the charge migration dynamics, can be recovered by the diagonal \ac{CASPT2} correction similar to benzene.

  \subsection{Ultrafast spin-flip dynamics}\label{subsec:spin-flip}
  
 Another type of dynamical process for which \RD\ is particularly suited is the dynamics in core-excited systems triggered by ultrashort X-ray pulses.
 For instance, the approach implemented within the \RD\ module has been used to study spin dynamics for excitation at the L-edge of transition metal complexes.~\cite{Wang_PRL_2017,Wang_MP_2017,Wang_PRA_2018,Kochetov_JCP_2020}
 We continue discussing these applications here, shifting the focus to methodological issues.
 The process under study can be understood as follows:
 After absorption of an X-ray photon, the localized core-hole is created.
 If the angular momentum of the core hole is non-zero, one can use the broad (ultrashort) pulse to create a superposition of pure spin states, which then evolve in time, resulting in spin-mixing or even spin-flip due to the strong SOC for core orbitals.
 In a sense, it is analogous to electron correlation-driven charge migration, but instead of oscillating hole population, one observes \ac{SOC}-driven spin oscillations.
 For the first-row transition metals, the \ac{SOC} constant for the $2p\rightarrow3d$ excitations is of the order of 10\,eV,~\cite{Kochetov_JCP_2020} giving a characteristic timescale of $\approx 400$\,(as).
 Given this timescale and relatively large masses of transition metal atoms and atoms in the first coordination sphere of the typical ligands, one can assume the approximation of clamped nuclei, inherent to \RD, to be particularly valid for this case.

 As mentioned in Sec.~\ref{subsec:charge_migration}, even using small and medium-sized active spaces often results in a large number of stationary basis states.
 Considering all of them to study dynamics can be connected with significant computational efforts or even be impossible.
 That is why the reduction of dimensionality may be critical. 
 {In cases like the charge migration in iodoacetylene, one can preselect basis states based on additional information available from the experiment or other \textit{a priori} considerations.
 For instance, in the case of \hcci, it was known from Ref.~\citenum{Kraus_S_2015} which states are mainly populated by the incoming light pulse that allowed us to substantially limit the number of basis \ac{SO} states from 12096 to numbers below 200, cf. Table~\ref{tab:charge_migration} and Fig.~\ref{fig:hcci_pes}(c).
 However, the knowledge about the initial state and characteristics of the excitation pulse can also help to \textit{a priori} reduce the computational effort in general cases.
 For example, for charge migration, one can decide in favor of some initial \ac{CSF} and knowing \ac{CI} Hamiltonian matrix preselect only those basis \acp{CSF} coupled to it by off-diagonal matrix elements directly, indirectly via single configuration, two configurations, and so on.
 Thus, one builds a kind of \ac{CI}-like hierarchy of the important states, which can be truncated according to accuracy and computational effort demands. 
 }

\begin{figure}
 	\includegraphics[width=0.4\textwidth]{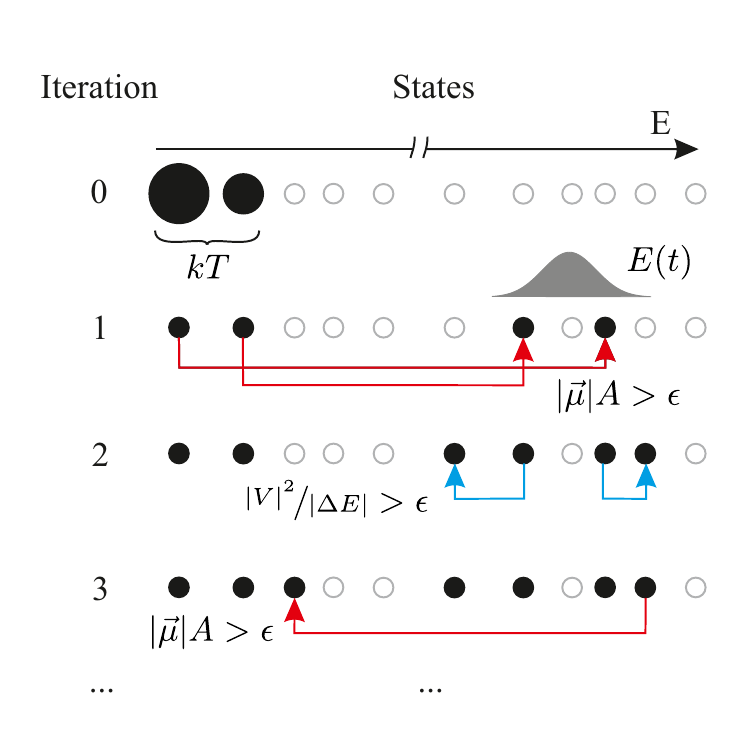}
 	\caption{\label{fig:selection} Iterative scheme to preselect states for dynamics. Iteration 0: thermally populated initial states. Iteration 1: light absorption within the energy window of the light pulse. Iteration 2: coupling between excited states. Iteration 3: population due to stimulated emission. Black circles denote participating states, open gray circles -- rejected states.}
 \end{figure}

 {In the case of spin dynamics triggered by an explicit light pulse, one should take into account the following: 
 i) the form of the initial state since there often exist several low-lying electronic states or spin-components of a multiplet which can be (thermally) populated; 
 ii) the excitation energy window due to bandwidth of the light pulse;
 iii) the magnitude of the transition dipole matrix elements which connect the initial basis states with those falling within the energy window;
 iv) the actual coupling matrix, e.g., $\mathbf{V}_{\rm SOC}$ which connects excited states and governs the dynamics.
 Accounting for i)-iv), one preselects first-rank participating states coupled via dipole transition and then the second-rank participating states coupled via \ac{SOC}.
 Note that if the pulse is strong enough, it may cause stimulated emission populating more states.
 Therefore, one would require iterative selection of participating states, depicted in Fig.~\ref{fig:selection}.
 In this case, the sorting into participating and spectator states is done according to a single threshold parameter $\epsilon$, where, provided the state $i$ is populated in the previous iteration, the state $j$ is rejected if $|\vec{\mu}_{ij}|A<\epsilon$ and $|V_{ij}|^2/|E_i-E_j|<\epsilon$, with $\vec{\mu}_{ij}$, $A$, $V_{ij}$, and $E_i$ being the transition dipole, field amplitude, \ac{SOC} matrix element, and basis-state energy, respectively. 
 }
 %
 
 Below we present the application of selective ultrafast spin dynamics at the L-edge for two iron complexes -- \FeHO\ and \FeCO. 

 
 \paragraph{Hexaaquairon (II)}
  \begin{figure*}
 	\includegraphics[width=0.75\textwidth]{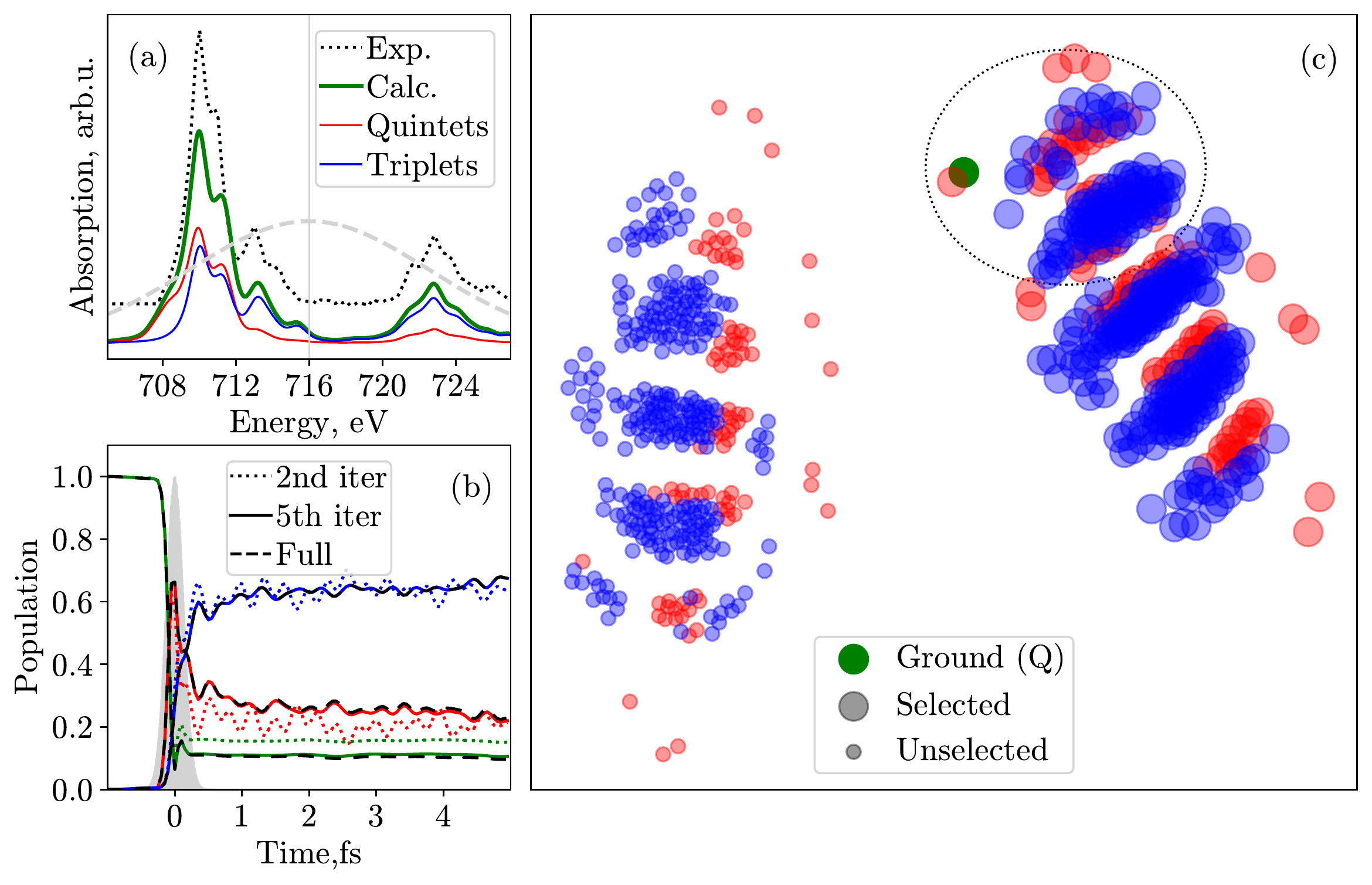}
 	\caption{\label{fig:feh2o6} Spin-dynamics after the L-edge excitation in the \FeHO complex: 
 		(a) Experimental and calculated \ac{XAS} with its decomposition in spin multiplicities. 
 		The dashed gray line shows the incoming pulse in the frequency domain. 
 		Pulse characteristics are $A=6$\,a.u., $\hbar\Omega=716$\,eV, $\sigma=0.2$\,fs.
 		(b) The time evolution of the population calculated including 128 and 378 preselected states (iterations 2 and 5, dotted and solid color lines, respectively) and using the basis of all 760 \ac{SF} states (dashed black lines). 
 		The filled gray curve shows the time envelope of the pulse.
 		(c) Force-directed graph~\cite{Fruchterman_SPE_1991, Hagberg_Pot7PiSC_2008} displaying connections between states due to transition dipole and \ac{SO} coupling. A circle represents each \ac{SF} state; large circles correspond to states participating in the dynamics. 
 		The color indicates spin multiplicity: red for quintets $S=2$, blue for triplets $S=1$, and green for the initially populated quintet state.
 	    Dashed circle displays the states selected after the second iteration; see text.}
 \end{figure*}
 \FeHO\ ion is one of the coordination complexes known to have a spin-flip after ultrashort X-ray pulse, i.e., to acquire a spin distinct from its ground state after X-ray excitation.~\cite{Wang_PRL_2017,Kochetov_JCP_2020}
 The computational scheme for the electronic structure in this work coincides with Ref.~\citenum{Kochetov_JCP_2020}.
 We employ the \ac{DKH} relativistic Hamiltonian,~\cite{Douglas_AP_1974} all-electron ANO-RCC-VTZP basis set, and \ac{RASSCF}/\ac{RASSI} level of theory.
 A reasonable active space with eight orbitals (three Fe $2p$ and five Fe $3d$) and 12 electrons was used, which resulted in 760 \ac{SO} basis states.
 The calculated static L$_{2,3}$-edge absorption spectrum (Fig.~\ref{fig:feh2o6}a) is in good agreement with the experiment.~\cite{Bokarev_PRL_2013}
 In this panel, one can see the Fourier-transformed excitation pulse envelope and the decomposition of the spectrum into spin-free contributions showing that in some parts, the contribution from the spin (triplet) other than the ground-state one (quintet) is prevailing. 
 The initial density matrix was constructed by populating the lowest \ac{SO} state, equivalent to zero temperature.
 Dynamics were triggered by the short pulse excitation with characteristics chosen to cover a wide range of valence-core $2p\rightarrow 3d$ excitations and make the ground state undergo substantial depletion up to 90\%, see Fig.~\ref{fig:feh2o6}(a).
 As displayed in panel (b) of Fig.~\ref{fig:feh2o6}, the initially populated quintet core-excited state mix with triplet states due to strong \ac{SOC} (12.8~eV constant) resulting essentially in a spin-flip.

 For visualization of connections between states due to transition dipole and \ac{SOC} matrix elements discussed above and used for the selection scheme, Fig.~\ref{fig:selection}, we use a force-directed drawing algorithm.~\cite{Fruchterman_SPE_1991, Hagberg_Pot7PiSC_2008}
   	The results are shown in Fig.~\ref{fig:feh2o6}(c).
   	Each node corresponds to one of the 760 \ac{SF} basis states, and the color encodes their multiplicity, i.e., red quintets and blue triplets; the initially populated state is green. 
   	The distances between nodes are optimized to minimize spring-like forces between them.
   	The force ($F_{ij}=-\kappa_{ij}\Delta x_{ij}$) between nodes $i$ and $j$ corresponds to the spring constant $\kappa_{ij}=c|({V}_{\rm SOC})_{ij}| + |d_{ij}|$, where $c$ is a factor governing the relative importance of the two couplings. 
   	It has been adjusted for visual clarity to illustrate the clustering of states.
   	The size of the nodes, in turn, denotes whether the state is involved in dynamics or stays mainly unpopulated.
    The states are separated into two main clusters, neither connected strongly by dipole transitions nor \ac{SOC}, whereas interaction is notable within the clusters. 
    It is natural to expect that half of the states from the left cluster are not participating in the dynamics.
 Indeed, for the threshold value $\epsilon = 27.2\,meV$, the iterative procedure described above (Fig.~\ref{fig:selection}) converged after five iterations. 
 The number of states selected at each iteration starting from 0 is 1, 8, 128, 240, 351, and 378.
 The states selected after iteration 2, i.e., the smallest reasonable basis where dipole and \ac{SOC} coupling are minimally accounted for, are denoted with the dashed ellipse in panel (c).
 The respective dynamics with 128 states shown in panel (b) with dotted lines demonstrate correct trends but are still different from the full one with 760 states (black dashed lines).
 However, after the fifth iteration, the dynamics with 378 are barely different from the full one.
 Thus, preselecting only a half of all states, which are marked with big nodes in Fig.~\ref{fig:feh2o6}c, produces the converged result.

 One can also analyze how states are added according to their spin magnetic quantum number $S_z$ and \ac{SOC} selection rules.
 It can be traced because the states additionally form microclusters according to the $S_z$ value within both larger clusters. 
 The initially populated state is the ground state with $S_z=-2$. 
 The quintet states with $S_z=-2$ were selected at the first iteration according to the dipole selection rule $\Delta S_z=0$.
 Both triplet and quintet states with $S_z=-1$ are also selected at the second iteration due to the \ac{SOC} selection rule $\Delta S_z=0,\pm 1$. 
 These two groups of states correspond to two subclusters encircled with the dashed line in Fig.~\ref{fig:feh2o6}(c).
 The further three iterations select consecutively states with $S_z=0,+1,+2$.


 \paragraph{Iron pentacarbonyl}
 \begin{figure*}
 	\includegraphics[width=0.75\textwidth]{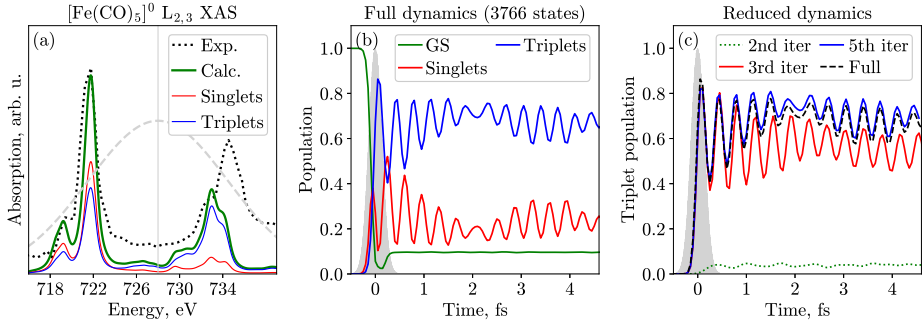}
 	\caption{\label{fig:feco5} Spin dynamics after the L-edge excitation of the \FeCO\ complex: 
 		(a) Experimental and calculated \ac{XAS} with its decomposition in spin multiplicities -- singlets (red line) and triplets (blue line). 
 		The dashed gray line shows the incoming pulse in the frequency domain; pulse characteristics are $A=6$\,a.u., $\hbar\Omega=728$\,eV, $\sigma=0.2$\,fs.
 		(b) Time evolution of the population singlet and triplet states calculated including all 3766 \ac{SF} states.
 		(c) Differences in the population of triplets (equivalent to triplet yield $\langle \hat{S}^2 \rangle /2 $) accounting for the total number of states as compared to the reduced dynamics, including only 111 (green, second iteration), 656 (red, third iteration), and 1882 (blue, fifth iteration) states.}
 \end{figure*}
 The \FeCO\ exhibits a significant spin-flip rate from the singlet to the triplet state manifold upon L-edge X-ray excitation~\cite{Kochetov_JCP_2020} and has also been used to test the preselection procedure.
 For this complex, active space consists of three $2p$ orbitals (RAS1, 1 hole is allowed) $3d\sigma$ ($a_1^{\prime}$), four $3d$ ($e^\prime$ and $e^{\prime\prime}$), and $3d\sigma^\ast$ ($a_1^{\prime\ast}$) (RAS2, full \ac{CI}) and four $\pi^{\ast}$ orbitals (RAS3, 1 electron is allowed), resulting in 13 orbitals with 14 active electrons;~\cite{Suljoti_ACIE_2013} other computational details coincide with \FeHO, see also Ref.~\citenum{Kochetov_JCP_2020}.
 This setup results in 3766 \ac{SO} states.
 Therefore, preselection of states, in this case, is more crucial for efficient computation than for \FeHO. 

 The absorption spectrum, full dynamics with 3766 states, and the triplet yield for the different number of selected states are displayed in Fig.~\ref{fig:feco5}.
 For \FeCO, the transition from singlet to triplet multiplicity is accompanied by strong Rabi-like oscillations.
 As was shown previously, not all states equally contribute due to different \ac{SOC} and transition dipole moment matrix elements.
 Similar to \FeHO, iron pentacarbonyl also shows clustering of states in two groups,~\cite{Kochetov_JCP_2020} but, in contrast, one also has subgroups due to substantially different transition dipole moments as $2p\rightarrow\pi^\ast$ transitions are more intense than the $2p\rightarrow3d$ ones.
 
 The threshold value of $\epsilon = 1.36$\,eV was applied for the preselection.
 As for \FeHO, the convergence to half of all states was also reached after five iterations.
 At each iteration starting from 0, the number of qualified states was 1, 9, 111, 656, 1599, and 1882, respectively.
 As shown in Fig.~\ref{fig:feco5}(c), the convergence to the full result is notably slower in this case.
 However, the dynamics with 1882 basis states almost quantitatively agree with the full dynamics.
 We observe that already at the third iteration with 656 states included, the dynamics are qualitatively reproduced, which is enough to describe such main features as the oscillation period and noticeable spin-flip rate.
 Finally, we note that the energetic distance between states plays an important role.
 Since for \FeCO\ the states lie much denser than for \FeHO, the threshold has to be selected about two orders of magnitude higher.
 It is also in accord with the larger transition dipole moments of the $2p\rightarrow\pi^\ast$ transitions compared to the $2p\rightarrow3d$ transitions.

\section{Conclusions and outlook}\label{sec:conclusion}
  
  In this article, we presented a program module \RD\ incorporated within the \molcas\ project.
  Its purpose is to study ultrafast electron dynamics on a level of complete or restricted active space \ac{CI} in the density-matrix formulation.
  Thus, it represents a straightforward extension of the stationary quantum chemistry available in the \molcas\ package to the time domain.
  Although the clamped nuclei approximation is inherent to the underlying theory, the effect of nuclear vibrations can still be taken into account in the form of harmonic vibrational heat-bath, ensuring the dissipation dynamics. 
  Thus, the methodology is particularly suited for studies of sub-few femtosecond electron dynamics when a system is excited far from conical intersections on the potential energy surface or when heavy atoms are involved.
  It can also be applied in cases when \ac{SOC} is important, staying, of course, within the limits of the applicability of the $LS$-coupling approximation. 
  Since the number of states belonging to different spin manifolds can be particularly large in the case of \ac{SOC}-mediated dynamics , the scheme for the preselection of the participating basis states is suggested.
  Therefore, the computational effort can be notably reduced.
  
   Several examples illustrate the possible applications of the methodology: computation of (non)-linear spectra, i.e., linear \ac{XAS} of \TiO\ and \ac{HHG} in H$_2$, charge migration in benzene and iodoacetylene, and the spin-dynamics in core-excited iron complexes.
  The density-matrix formulation of the \ac{CI} problem not only allows for the treatment of energy and phase relaxation but also offers a convenient way to incorporate (auto)ionization which will be the subject of our future research.
  
  
%
  

\section*{Acknowledgements}
	Financial support from the Deutsche Forschungsgemeinschaft Grant No. BO 4915/1-1 is gratefully acknowledged.

\bibliographystyle{ieeetr}
\bibliography{RhoDyn}
 
\end{document}